\documentclass[11pt]{article}
\usepackage{moriond,epsfig,graphicx,natbib,amssymb,psfig}

\def\be{\begin{equation}}
\def\ee{\end{equation}}
\def\bea{\begin{eqnarray}}
\def\eea{\end{eqnarray}}

\def \xmm {\hbox{\it XMM-Newton }}
\def \chandra {\hbox{\it Chandra }}

\def\kT {{\rm k}T}

\def\keV {\rm keV}

\def\rv {R_{200}}

\def\rfive {R_{500}}

\def \s01 {S_{0.1}}

\def \mt {$M$--$T$}
\def \lxt {$L_X$--$T$}

\newcommand{\propsim}{\lower 3pt \hbox{$\, \buildrel {\textstyle
      \propto}\over {\textstyle \sim}\,$}}

\begin{document}
\vspace*{4cm}
\title{X-RAY OBSERVATIONS OF THE MASS AND ENTROPY DISTRIBUTIONS IN NEARBY GALAXY CLUSTERS}

\author{ G.W. PRATT }

\address{MPE Garching, Giessenbachstra{\ss}e, 85748 Garching, Germany}

\maketitle\abstracts{I review some important aspects of the structural
and statistical properties of the nearby X-ray galaxy cluster
population, discussing the new constraints on mass
profiles, the mass-temperature relation, and the entropy of
the intracluster medium which have become available from recent X-ray
observations.}

\section{Introduction}
\label{sec:intro}

Massive systems of galaxies, gas and dark matter, located at the
intersection of tenuous, large-scale filaments, galaxy clusters are
the nodes of the highly structured Universe we see today. Formed from
the collapse of initial density fluctuations, clusters are built
hierarchically and constitute an evolving population. The evolution of
structure in the Universe -- and thus that of clusters -- is strongly
dependent on the cosmology, so that the statistical properties of the
cluster population are a test both of cosmology and of structure
formation theory itself. 

While clusters were initially discovered as overdensities of galaxies
on optical plates, our present understanding of their composition
suggests that the galaxies themselves constitute only $\sim 5$ per
cent of the total mass, with most of the mass ($\sim 80$ per cent) in
the dark matter component. The dominant baryonic component is the hot,
X-ray emitting intracluster medium (ICM), a rarefied plasma at several
million K, reflecting the great depth of the potential well in which
it lies. We are currently in the privileged position of having several
world-class X-ray observatories with which to examine the structural
and scaling properties of the ICM in galaxy clusters. The arcsecond
spatial resolution of {\it Chandra\/} is complemented by the grasp of
{\it XMM-Newton}, allowing spatially-resolved observations to be performed with
unprecedented precision\footnote{In addition, the recently-launched
  {\it Suzaku}  
  X-ray satellite has a much lower instrumental background, holding
  the promise of spatially resolved observations to larger
  cluster-centric distances.}.

In this paper I begin with a brief review of the expectations for
structure and scaling from the starting point of simple gravitational
structure formation.  I then compare these simple expectations with
recent observations of nearby clusters from {\it XMM-Newton} and {\it
Chandra}. I first show how X-ray observations can be used to place
strong constraints on the dark matter component, and then, passing via
the $M$--$T$ relation, I take a look at how measurements of the
entropy have cast light on the effect of
non-gravitational processes, such as galaxy formation and cooling, on
the X-ray properties of the cluster population.

\section{Expectations for structure and scaling}
\label{sec:scaling}

It is useful to have a baseline prediction for the X-ray properties of
the cluster population. Cluster formation is driven by the
gravitational collapse of the dominant dark matter component. The ICM
collapses with the dark matter, so that, to a first approximation,  
the gas properties should be determined entirely by the processes
involved in gravitational collapse, i.e., shock heating and
compression. The baseline model rests on several simple assumptions
\citep[e.g.,][]{eke}: 

\begin{itemize}\setlength{\itemsep}{0.25ex}

\item The internal structures of clusters of different mass are
  similar.

\item All clusters identified at a given redshift correspond to a
  given characteristic density, and the characteristic density scales
  with the critical density of the Universe, i.e.,

\begin{equation}
\frac{GM_{200}}{R_{200}^3} = \langle \rho \rangle = \frac{4\pi}{3} \delta \rho_c (z) ; \delta \sim 200.
\end{equation}

\item The ICM evolves in the gravitational potential of the dark
  matter; the gas mass fraction is thus constant,

\begin{equation}
f_{\rm gas} = \frac{M_{\rm gas}}{M_{200}}.
\end{equation}

\item Time-averaged, the ICM is in approximate hydrostatic
  equilibrium, allowing application of the Virial Theorem, giving:

\begin{equation}
\frac{G \mu m_p M_{200}}{2 R_{200}} = \beta_T {\rm k}T
\end{equation}

\noindent where $T$ is the mean cluster X-ray temperature and
$\beta_T$ is a normalisation constant dependent on internal structure.

\end{itemize}

\noindent The above equations imply that power-law relations exist
between various X-ray properties, $Q$, and the mass $M$ (or a mass
proxy) and the redshift, $z$, such that $Q \propto A(z)
M^{\alpha}$. The evolution factor, $A(z)$, is due to the evolution of
the mean density, which varies with the critical density, $\rho_c(z)
\propto h^2 (z)$. For example, the gas and total mass scale as $M_{\rm
gas} \propto M_{200} \propto h^{-1}(z) T^{3/2}$, and, assuming
bremsstrahlung emission, the X-ray luminosity scales as $L_X \propto
h(z) T^2$. The assumption of structural similarity further implies
that, once scaled appropriately, the radial profiles of various
cluster quantities (e.g., gas density $\rho$, entropy $S$, temperature
$T$, total mass $M$) are similar.

Clusters created in hydrodynamical simulations of purely gravitational
structure formation are structurally similar and follow these scaling
relations \citep[e.g.,][]{bn}, showing that the assumptions underlying
this simple spherical baseline model are applicable in a hierarchical
cosmological context\footnote{Note that there is a certain amount of
  intrinsic scatter in the relations, which is related to cluster
  dynamics.}. In comparison with 
observations, the study of 
both the correlations between global properties (the scaling
relations) and the structural properties (radially-averaged profiles
of various quantities) is essential for the best understanding of
cluster formation and evolution.


\section{Dark matter constraints} 

\subsection{Mass profiles}

An inescapable prediction from more than a decade's worth of numerical
simulations is the existence in haloes of a semi-universal cusped
dark matter density profile. The most well-known variant is the
Navarro-Frenk-White \citep[][NFW]{nfw} profile, $\rho_{DM}(r) \propto
[(r/r_s) (1 + r/r_s)^2]^{-1}$, where the concentration parameter,
$c_{\delta} = R_{\delta}/r_s$, is a measure of the halo
concentration. The profile is not strictly universal since there is a
weak dependence of $c_{\delta}$ on mass, $M_{\delta}$, due to the fact that
smaller mass haloes generally form earlier, when the Universe was
denser \citep{bullock,dolag}. Although the total mass density profile
can be measured with both X-ray and 
optical observations, in the latter case, stacking is often required to
overcome the paucity of data on individual objects if quantitative
measurements of the mass profile are to be obtained \citep[e.g.,][]{biv}. 
In contrast, while X-ray observations allow measurements of individual
mass density profiles, the method relies on the assumption that the
gas is in hydrostatic equilibrium. 

\begin{figure}
\begin{centering}
\includegraphics[scale=1.,angle=0,keepaspectratio,width=0.55\columnwidth]{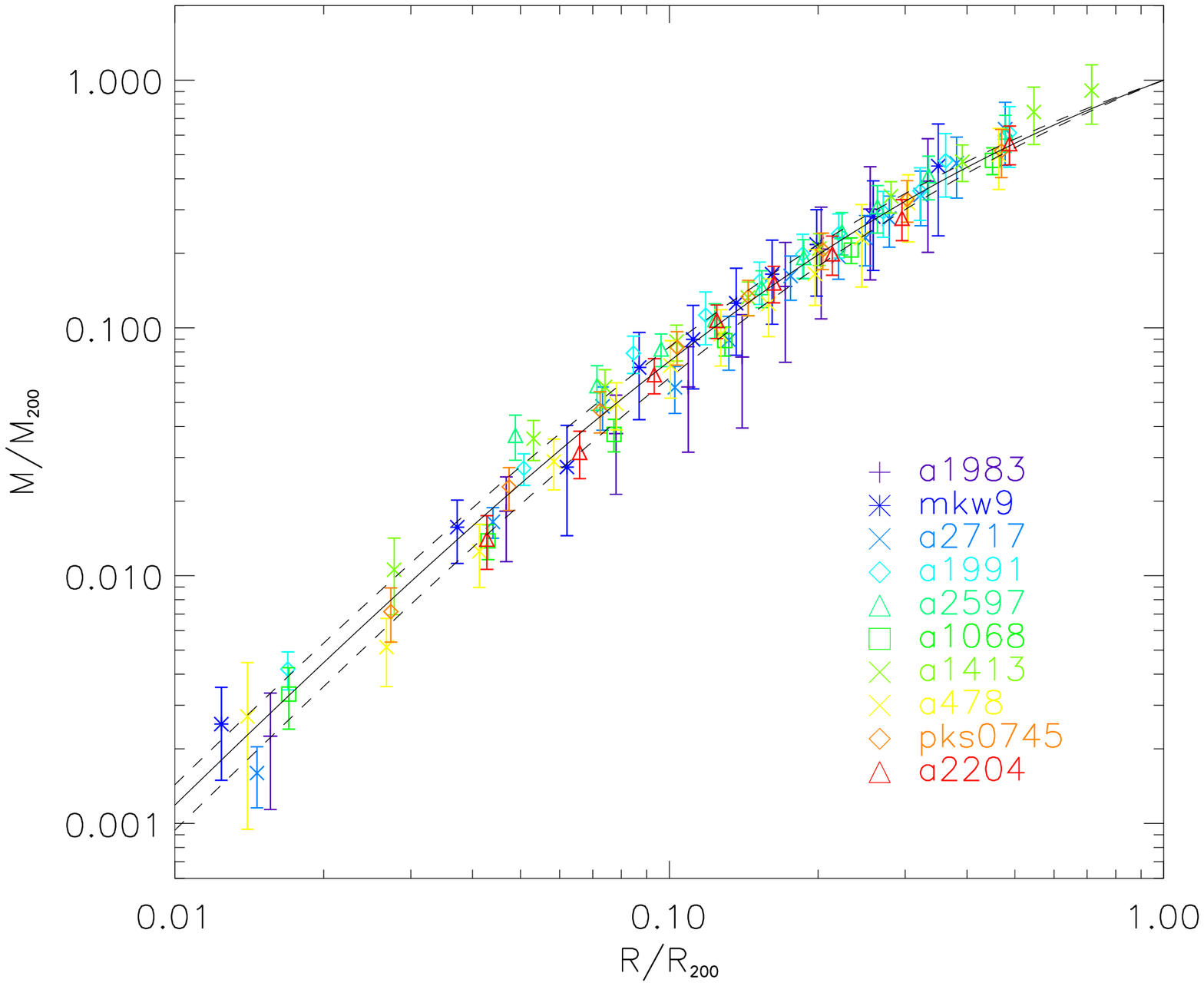}
\hfill
\includegraphics[scale=1.,angle=0,keepaspectratio,width=0.44\columnwidth]{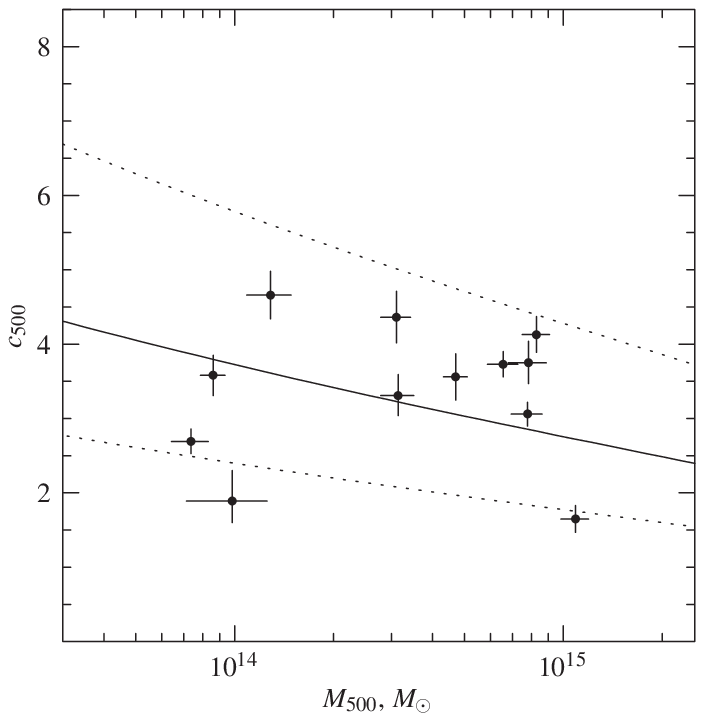}
\caption{{\footnotesize {\bf Left:} Integrated mass profiles of 10
    clusters in the temperature range 2 - 9 keV, observed with
    \xmm \citep[from][]{point05}. For each cluster, the mass is scaled to
    $M_{200}$ and the radius 
    to $\rv$. The black line corresponds to the mean scaled
    best-fitting NFW model;, the dashed lines the associated standard
    deviation. {\bf Right:} Concentration parameter vs mass relation
    for NFW model fits to the total density profiles of 12 clusters in
    the temperature range 0.7 - 9 keV, observed with \chandra
    \citep[from][]{vikh06}. The solid line shows the average concentration of
    CDM haloes from simulations; the dotted lines show the associated
    $2\sigma$ scatter. }}\label{fig:mprofs} 
\end{centering}
\end{figure}

 X-ray observational confirmation
of the existence of a cusped profile on 
cluster scales started to become available with {\it ASCA\/}
observations \citep[][]{mark2}, although the spatial resolution of
these observations, in particular of the temperature profiles, was not
optimal. More recent observations with \xmm 
and \chandra have allowed the mass profiles of relaxed clusters to be
probed from the very core regions \citep[$\sim 0.01 \rv$,
e.g.,][]{lewis} right out to $0.7-0.8\,\rv$ \citep[${\rm i.e., }\ \sim
\rfive$; ][]{pa02,point04,vikh06,vf06}. All of these observations point
to the existence of a quasi-universal cusped profile as predicted from
CDM simulations. There is no evidence
for a core in even the highest-resolution observations \citep{lewis},
effectively ruling out the possibility of self-interacting dark
matter on cluster scales.  

\subsection{Concentration parameters}

Observations of small samples of clusters and galaxy groups have also
allowed the universality of mass and total density profiles to be
probed \citep{pa05,point05,vikh06,hum}. The left-hand panel of
Fig.~\ref{fig:mprofs} shows the integrated mass profiles of 10
clusters in the temperature range 2 - 9 keV, scaled by the virial mass
$M_{200}$ and radius $\rv$. The mass profiles can all be adequately
fitted with NFW models and are clearly similar once scaled in this
manner. The right-hand panel of Fig.~\ref{fig:mprofs} shows the
measured concentration parameters from fits to the total density
profile of 12 clusters in the temperature range 0.7-9 keV
\citep{vikh06}, compared to the theoretical predictions of
\citet{dolag}. The observed concentration parameters are in good
agreement, both in absolute value and dispersion, with the theoretical
predictions. Similar results have been found by \citet{point05} for
another sample of clusters, and by \citet{hum} for a sample of early
type galaxies and galaxy
groups. In the latter case, it was found that inclusion of a stellar
mass component was necessary to give acceptable fits to the data,
since this component starts to dominate the mass budget in the very
inner regions (see
also \citealt{lm}).  Such results are encouraging and have allowed
constraints to be put on the nature of the dark matter itself on
cluster scales, suggesting that our understanding of the dark matter
collapse is correct down to the scale of individual giant galaxies.

Finally, it should be pointed out that all of the above results were
obtained only for morphologically relaxed systems. Attempts to
model the total mass density profiles of less relaxed systems with the
NFW profile have resulted in poor fits \citep[][]{bel,pbf}. This suggests
that in these cases either (i) the gas in not in hydrostatic equilibrium,
or (ii) that the underlying DM density profile is not well described by an
NFW model (the original NFW model was proposed specifically for
systems with a relatively high degree of relaxation).

\section{The $M$--$T$ relation}

Models of structure formation predict the space density, distribution
and physical properties of the cluster population as a function of mass
and redshift, so that, from the theoretical point of view, the mass of
a cluster is its most fundamental property. However, X-ray cluster
surveys yield the cluster space density and distribution only in terms of
X-ray observables such as the luminosity, $L_X$, or the temperature $T$.
Scaling relations linking the mass to the X-ray observables allow the
full use of X-ray cluster survey information in the statistical sense.
Knowledge of mass observable relations and the associated scatter
about these relations is thus crucial in the use of
clusters to probe cosmology. 

Assuming hydrostatic equilibrium and spherical symmetry, the mass of a
cluster can be recovered from spatially-resolved X-ray observations of
the density and temperature profiles.
The \mt\ relation is one of the fundamental relations for linking the
observed gas properties with theory, since the temperature is expected
to be closely related to the mass via the Virial theorem. It is also
of topical interest, 
since the normalisation of this relation has a direct bearing on the
measurement of the present-day value of $\sigma_8$ using cluster
abundance measurements, and, until the WMAP 3-year data, cluster
measurements of $\sigma_8$ were thought to be unusually low.
Samples from {\it
  ROSAT\/} and {\it ASCA}, though large, either assumed that clusters
were isothermal \citep[][]{horner,xu,cms} or had relatively poor
temperature profile resolution and/or relied strongly on
extrapolation to derive the virial mass \citep[][]{neva,fin,sand}. 
As a result, historically, there has been little consensus on either the
slope or the normalisation of the \mt\ relation.

\begin{figure}
\begin{centering}
\includegraphics[scale=0.1,angle=0,keepaspectratio,width=0.495\columnwidth]{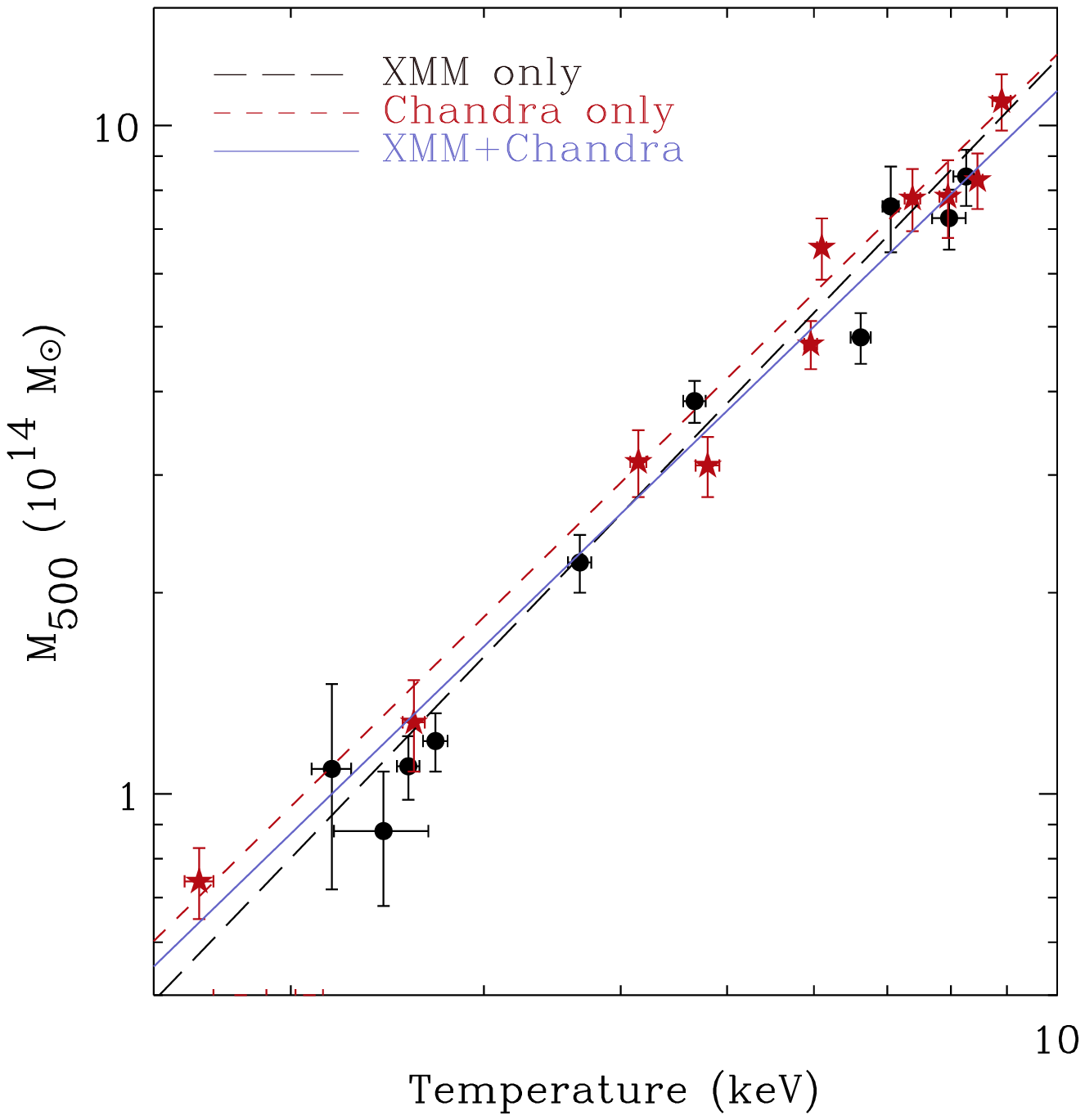}
\hfill
\includegraphics[scale=0.1,angle=0,keepaspectratio,width=0.495\columnwidth]{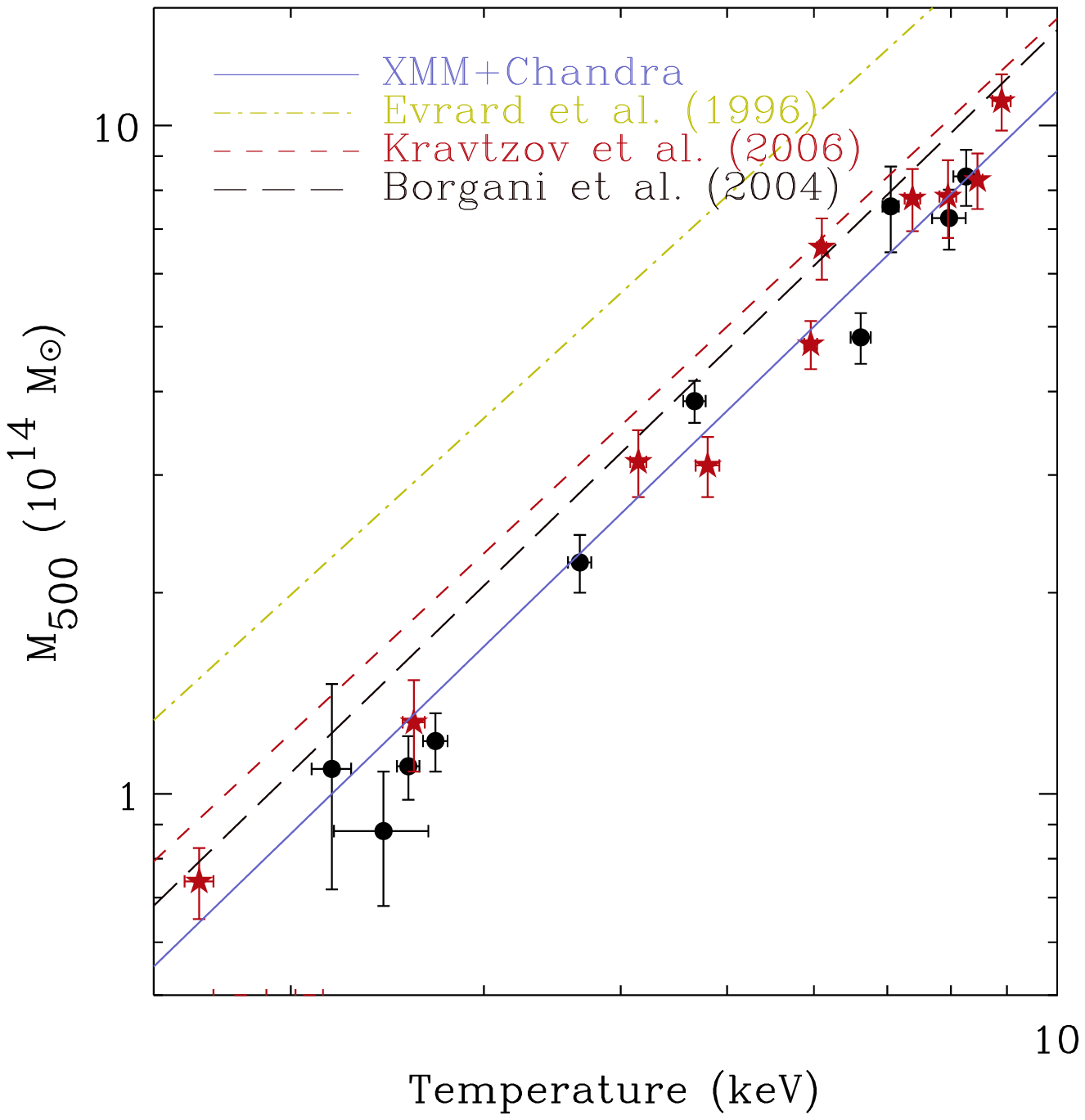}
\end{centering}
\caption{\footnotesize {{\bf Left:} The X-ray mass-temperature
    relation from observations obtained with \xmm \citep[][circles]{app} and
    \chandra \citep[][stars]{vikh06}. Some clusters were observed with
    both satellites, and are plotted twice. The long-dashed line is the
    best-fitting relation to the \xmm\ data points, the dashed
    line is the best fitting relation to the \chandra\ points, and
    the solid lines is the best fitting relation to all data
    points. See text for details. {\bf Right:} Comparison with
    results from several numerical simulations. The inclusion of more
    realistic physical processes in more recent simulations has
    improved the agreement between 
    observation and theory.}}\label{fig:mt}

\end{figure}

The situation is much improved with recent results from \xmm\ and {\it
Chandra}. The left-hand panel of Figure~\ref{fig:mt} shows the
mass-temperature data from two cluster
samples observed independently with \xmm \citep{app} and \chandra
\citep{vikh06}\footnote{Similar results were found at $R_{2500}$ for a
  smaller 
sample of 6 systems by \citet{allen}}. The various lines show the 
results of fitting a power-law 
function of the form $h(z) M = A \times (\kT /5\ \keV)^\alpha$ to
different data sets. The fits to the individual \xmm and \chandra data
sets give results which are entirely consistent within the $1\sigma$
errors for both slope and normalisation. A BCES fit to the total data
set yields the relation   

\begin{equation} 
\log{[h(z) M_{500} / 10^{14}\,M_{\odot}]} = (0.57\pm0.02) + (1.59\pm0.08) \log{[\kT / 5\
  \keV]}. 
\end{equation}

\noindent  The slope is thus very close to the expectations from the
simple scaling analysis in Sect.~\ref{sec:scaling} above. 

A second historical point of contention is the disagreement between
the observed \mt\ normalisation and that derived from numerical
simulations. Now that there is good agreement between measures of the
observed normalisation, the time is ripe for a new comparison with
simulations. The right-hand panel of Figure~\ref{fig:mt} shows the
data and best fitting power law described above, compared to the
relations found in the simulations of \citet{emn}, which only include
gravitational processes, and the more recent simulations of
\citet{borg} and \citet{krav}, which include additional physical
processes such as radiative cooling and feedback from supernovae. The
disagreement 
with the gravitation-only simulations is well documented and for the
present data set amounts to a difference in normalisation of 40 per
cent at 5 keV. The agreement has clearly improved with the
introduction of more realistic physical processes into the
simulations.

The remaining discrepancy is likely due to a combination of two
effects. Firstly, it is unclear how simplifying assumptions of the
X-ray analysis have an impact on the final mass determination. In
particular, the validity of the assumption of hydrostatic equilibrium
has often been called into question. Observations of Coma suggest a lower
limit of $\sim 10$ per cent of the total ICM pressure in turbulent
form \citep{ps}. However, Coma is a merging cluster and at present no
observational constraints exist for the amount of non-thermal pressure
support in clusters which observers would call `relaxed'. Secondly,
the value of the $M$--$T$ normalisation can vary by up to 50 per cent
depending on the simulation \citep{app,henry}, and the normalisation
has been creeping downwards over time as more complex physics has been
incorporated into the simulations. Current simulations are still
unable to reproduce not only the observed scaling relations (e.g., the
$L_X$--$T$ relation), but also the structural properties of clusters
(e.g., temperature profiles in the central regions), clearly pointing
to the possibility that certain physical processes may be
missing. Below I discuss how X-ray observations can give an idea of
what other physical processes may be in play.

Finally, it should be noted that the above
results were obtained for morphologically relaxed cluster
samples only. Since the precision on the estimation of cosmological
parameter estimation depends on the precision to which the scaling
relations are known \citep[e.g.,][]{henry}, accurate estimates of the
dispersion around the 
best fitting relations are needed {\it for the whole cluster
population}. This is true even for the case of self-calibration of
extremely large samples. The challenge now is to derive accurate
mass-observable scaling relations for {\it representative} cluster
samples. Since the X-ray method is clearly invalid for clusters which
are not in hydrostatic equilibrium, this will require
inter-calibration of the available mass estimators (X-ray, optical,
lensing), preferably using both observed and simulated data sets.  

\section{Gas physics}
\label{sec:gasp}

X-ray data have been telling us for nearly two decades that the simple
scaling relations outlined in Sect.~\ref{sec:scaling} do not describe
the observed cluster population. The most well-studied example is that
of the \lxt\ relation, which has been found by many studies to be $L_X
\propto T^3$ \citep[][]{es,ae,mark,op}, corresponding to a
suppression of X-ray luminosity in poor systems relative to simple
expectations. This suggests that physical processes other than gravity
alone are affecting the properties of the ICM, and that poorer (i.e.,
lower temperature) systems are proportionally more affected. These
non-gravitational processes are likely linked to radiative cooling and
to heating processes associated with galaxy formation.

\begin{figure}
\begin{center}
\includegraphics[scale=0.1,angle=0,keepaspectratio,width=0.515\columnwidth]{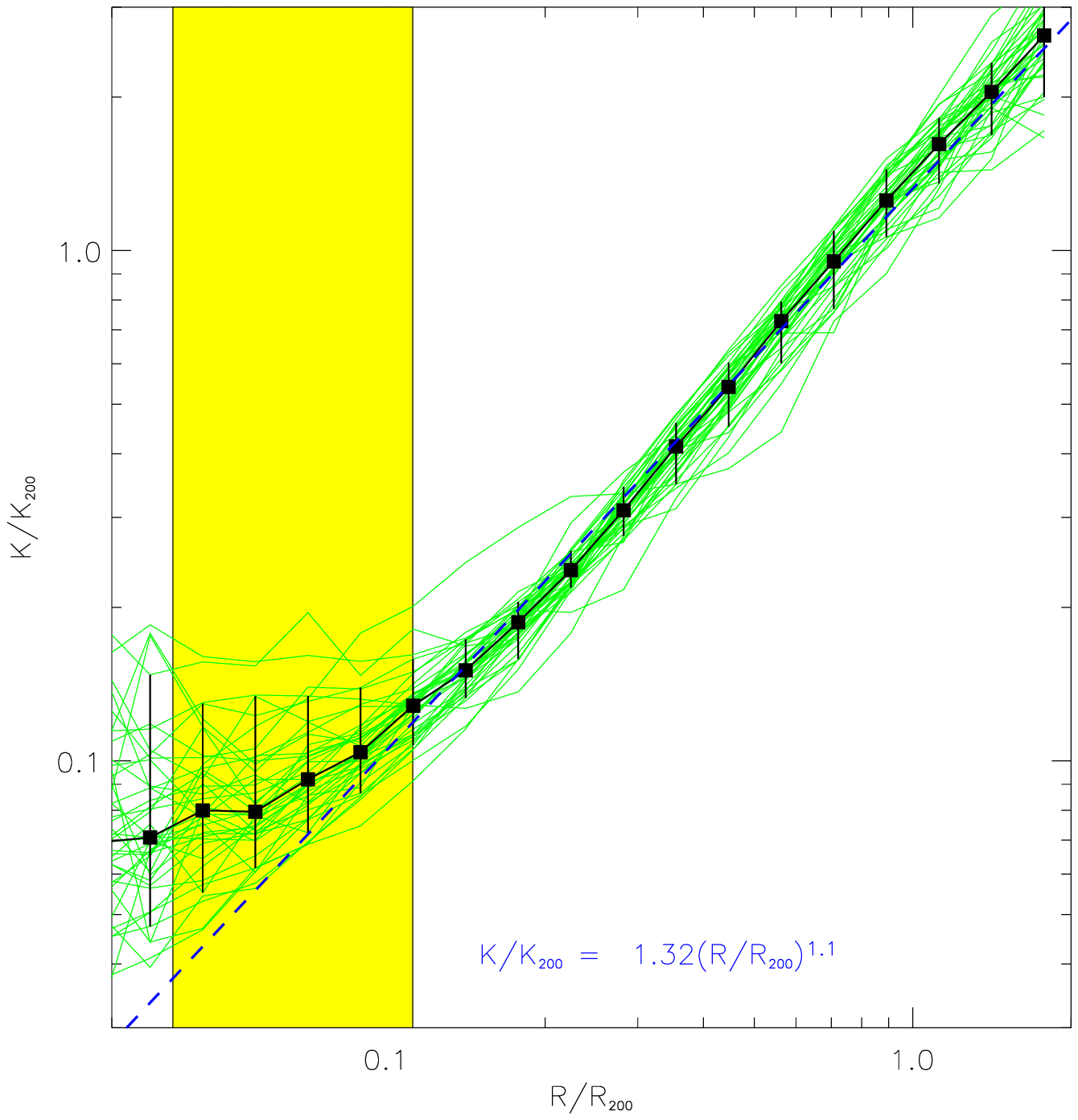}
\hfill
\includegraphics[scale=0.1,angle=0,keepaspectratio,width=0.475\columnwidth]{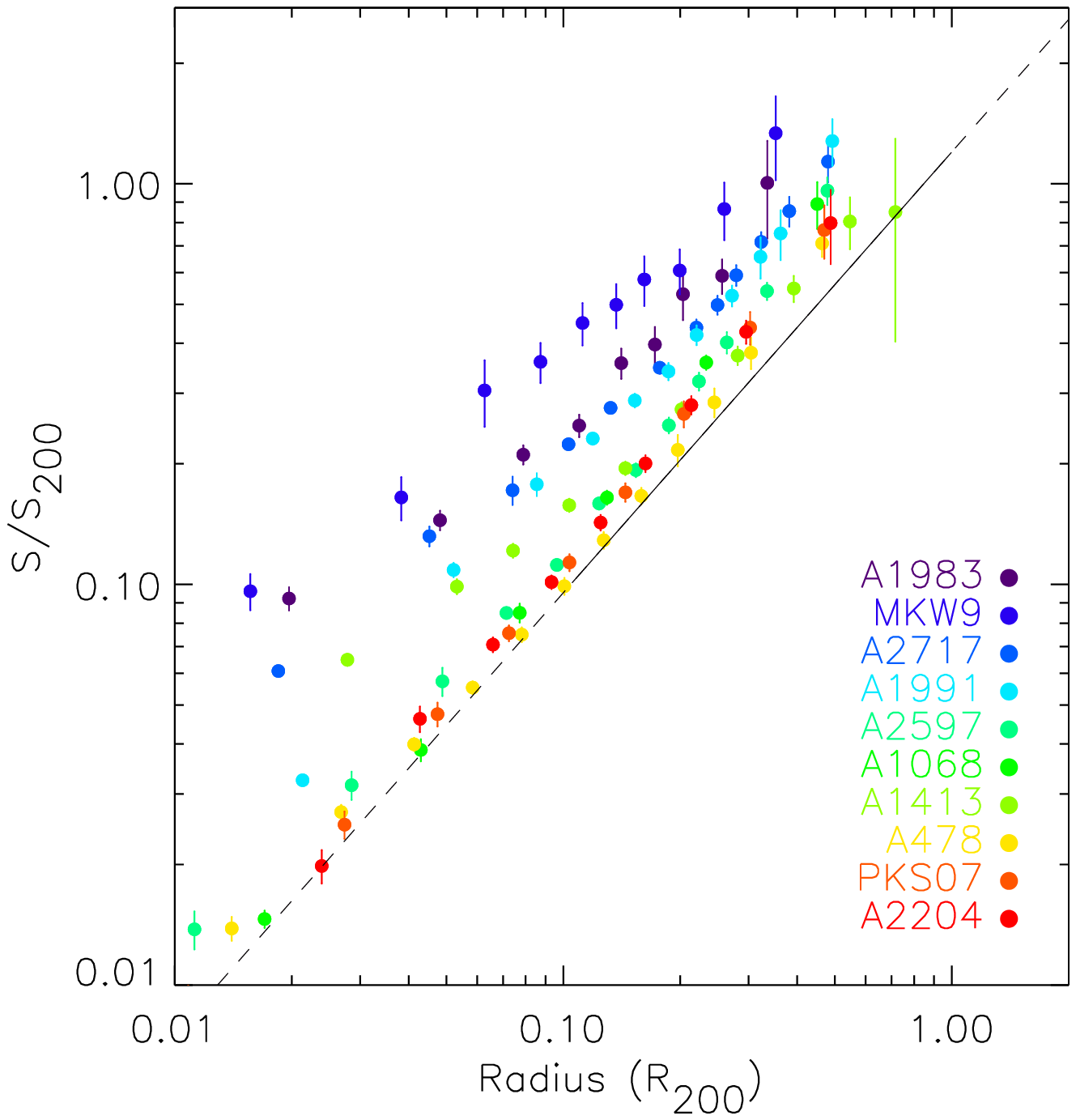}
\caption{\footnotesize {{\bf Left:} Entropy profiles of 40 clusters
    spanning the mass range $0.2 - 8 \times 10^{14}
    h^{-1}\, M_\odot$ from simulations including only gravitational
    processes \citep[from][]{vkb}. The profiles have been scaled using the
    characteristic entropy of the halo $K_{200}$ and the virial radius
    $\rv$ (see text for details). 
    {\bf Right:} Entropy profiles of 10 clusters spanning the
    mass range $1 - 12 \times 10^{14} h_{70}^{-1}\, M_\odot$ ($\sim 2$ [blue] -
    $8\, \keV$ [red]), observed with {\it XMM-Newton} \citep[from][]{pap}. The
    profiles have been 
    scaled in exactly the same manner. The solid line is the
    best-fitting power law description of the radial dependence of the
    entropy for the simulated clusters.}}\label{fig:ent}
\end{center}
\end{figure}

The entropy affords the most direct way of investigating the properties
of the ICM gas. The entropy\footnote{Here I follow the established
  convention and refer to 
  the quantity $S = K = \kT n_e^{-2/3}$ as the entropy. This is
  related to the true thermodynamic entropy via $s = \ln{K}^{3/2} +$
  constant.} is generated in shocks as the gas falls into the deep
potential of the cluster halo. Entropy is conserved in any adiabatic
rearrangement of the gas, and the gas will always rearrange itself so
that the entropy increases outwards. This property is illustrated in the
left-hand panel of Fig.~\ref{fig:ent}, which shows the
entropy profiles of 40 clusters spanning a mass range of a factor of ten,
from simulations including only gravitational
processes \citep{vkb}. The profiles have been scaled by the
characteristic entropy of the halo

\begin{equation}
K_{200} = \frac{1}{2} \left[ \frac{2 \pi}{15} \frac{G^2 M_{200}}{f_b H(z)}
  \right]^{2/3},
\end{equation}

\noindent where $f_b$ is the baryon fraction, and the radii have been
scaled to the virial radius $\rv$. It can be seen that the profiles
all coincide, and that, outside the central regions (where
gravitational softening becomes important in the simulations), a
power-law model with $K \propto R^{1.1}$ is an adequate description of
the trend with radius. In the right-hand panel of
Fig.~\ref{fig:ent}, I show the entropy profiles of ten
morphologically relaxed clusters spanning the temperature range 2-10
keV ($\sim 10^{14} - 10^{15}\, M_\odot$), observed with \xmm
\citep{pap}. The clusters are colour-coded so that poor (cool, low
mass) clusters are blue and rich (hot, high mass) clusters are
red. They have been scaled in exactly the same manner as the simulated
profiles, assuming $f_b = 0.14\ (\Omega_b h^2 =0.02$ and
$\Omega_m=0.3)$, and the solid line denotes the best-fitting power-law
entropy-radius relation to the scaled simulated clusters. It can be
seen that, while the profiles of rich clusters are in good agreement
(both in slope and in normalisation) with the prediction from
gravitational entropy generation, poor clusters have a systematically
higher entropy throughout the ICM.

\begin{figure}
\begin{center}
\includegraphics[scale=0.1,angle=0,keepaspectratio,width=0.525\columnwidth]{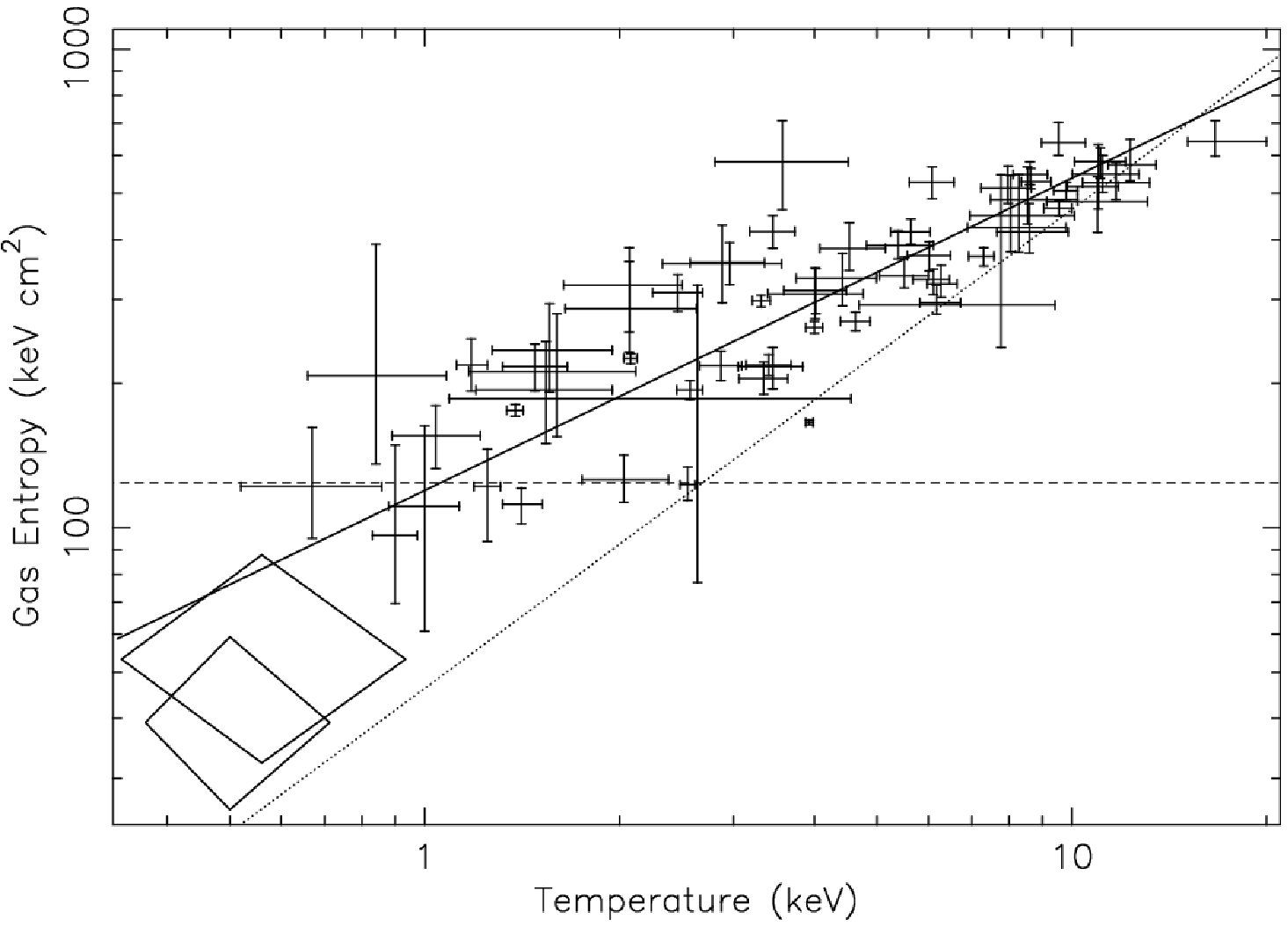}
\hfill
\includegraphics[scale=0.1,angle=0,keepaspectratio,width=0.45\columnwidth]{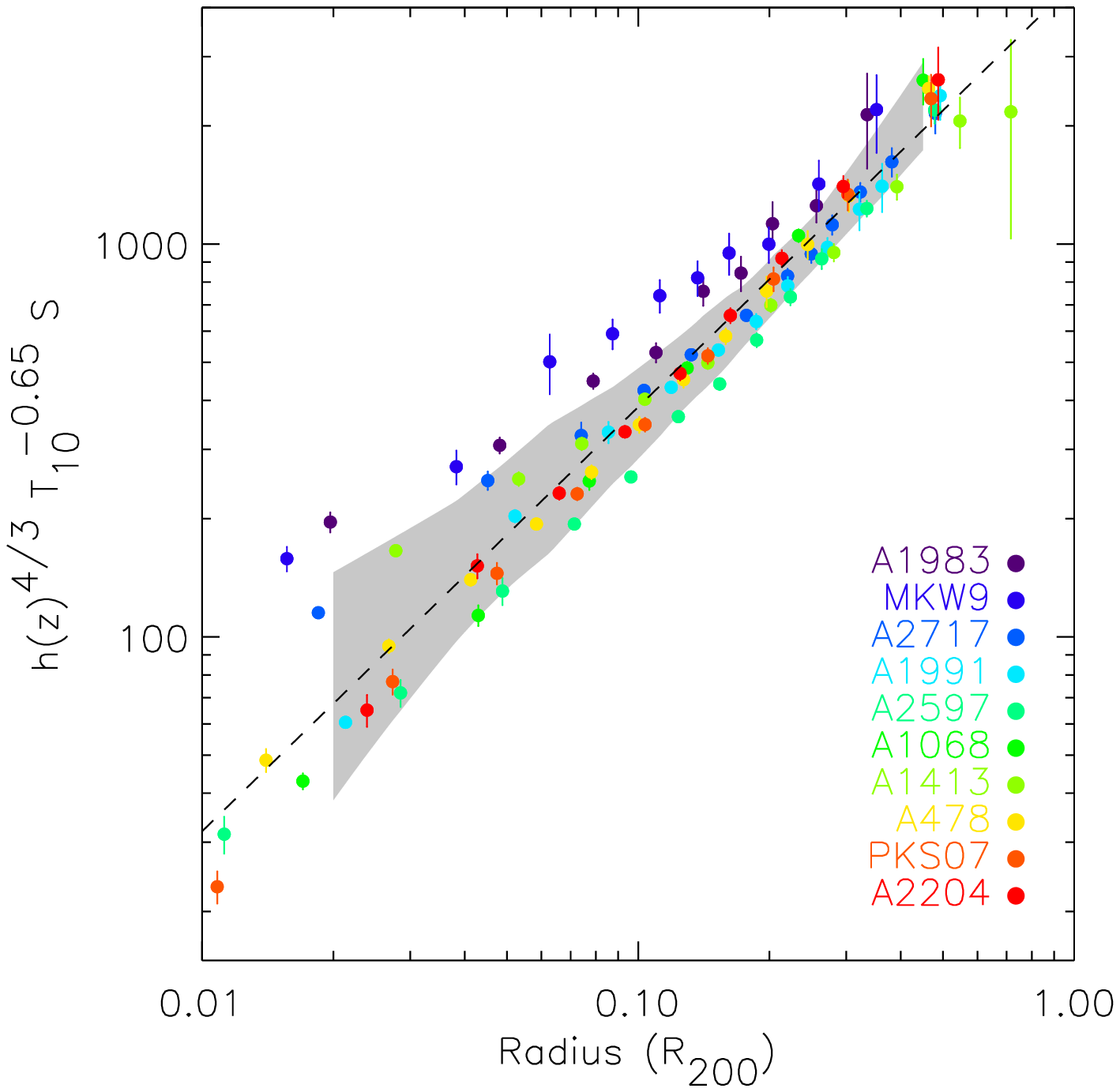}
\caption{\footnotesize {{\bf Left:} Entropy measured at $0.1\,\rv$ vs
    system temperature for 66 systems \citep[from][]{psf}. The dotted
    line shows the relation expected from pure gravitational structure
    formation ($S \propto T$). The solid line is the best-fitting
    power law relation: $S \propto T^{0.65}$. {\bf Right:} Entropy
    profiles of 10 clusters spanning the 
    mass range $1 - 12 \times 10^{14} h_{70}^{-1}\, M_\odot$ ($\sim 2$ [blue] -
    $8\, \keV$ [red]), observed with {\it XMM-Newton}
    \citep[from][]{pap}. The profiles have been 
    scaled using the best-fitting empirical entropy scaling. 
    }}\label{fig:ent2} 
\end{center}
\end{figure}

We can get an idea of the dependence of the offset with temperature (or
mass) by looking at the entropy measured at a certain fraction of the
virial radius. The entropy-temperature $S$--$T$ (or $K$-$T$) relation
was extensively investigated using data from the previous generation
of X-ray satellites. In the simple baseline model outlined in
Sect.~\ref{sec:scaling}, the gas mass fraction is constant, implying a
constant gas density and thus a simple linear scaling of the entropy
with temperature $S \propto h(z)^{-4/3} T$, or mass $S \propto
h(z)^{-2/3} M^{2/3}$. Initial measurements of the entropy at 
$0.1\,\rv$ \citep{pcn} showed a levelling-off of the $S$--$T$ relation
towards 
lower temperatures, which was subsequently interpreted as a
limiting value to the central entropy \citep{lpc}. However, deeper
investigation with larger samples observed with {\it ROSAT\/} and {\it
  ASCA} \citep{psf}, and smaller samples observed with the newest
generation of satellites \citep{pap} have shown that there is in fact
a continuous power-law relationship between the entropy at a fixed
fraction of the virial radius and the temperature, such that $S
\propto T^{0.65}$ (see the left-hand panel of
Fig.~\ref{fig:ent2}). This slope is in agreement with that expected
from the observed scaling of 
X-ray luminosity with temperature. The equivalent entropy-mass relation is $S
\propto M^{0.36}$, again shallower than expectations from simple
gravitational collapse \citep{pap}.

High resolution entropy profiles can give essential insights into the
physical processes at play. The right-hand panel of Fig.~\ref{fig:ent2}
shows the entropy profiles of the same ten clusters from
Fig~\ref{fig:ent}, this time rescaled by the observed 
empirically-determined entropy-temperature relation \citep[$S \propto
T^{0.65}$; see also][]{piff}. Two important points are evident from this
exercise. Firstly, outside 
the core regions, the entropy profiles show a remarkable degree of
similarity, and the radial dependence of entropy is well described by
the relation $S \propsim R^{1.1}$, as expected from standard
gravitational collapse (Sect.~\ref{sec:scaling}), and as is seen in
gravitation-only simulations \citep{vkb}. Thus non-gravitational
processes appear to alter the power-law scaling of the
entropy-temperature relation without breaking structural similarity,
at least for clusters of this mass range and above (results for the
group regime $\lesssim 2\, \keV$ are less conclusive, see
\citealt{mahd}). It is still 
unclear what mechanism is causing this effect. One proposal is that
smoothing of the gas density due to preheating in filaments and/or
infalling groups may boost entropy production at the accretion shock
\citep{voit03,psf}. However, numerical simulations run specifically to
test this effect seem to suggest that the boosting, though present,
appears to be substantially diminished in the presence of radiative
cooling \citep{borg05}. The same simulations suggest that feedback
from supernovae is too localised to have a significant effect on
smoothing of the accreting gas. Another possibility is that activity
from the central AGN produces the extra heating, although the observed
normalisations require than the AGN affects the entropy distribution
in a non-catastrophic manner at least out to $R_{1000} (\sim
0.5\,\rv)$ and it is not clear why and how this would produce the
observed power-law entropy-temperature dependence.

Secondly, inside the core regions the dispersion increases dramatically,
constituting an effective breaking of similarity. Three probable
mechanisms for breaking of similarity in the core are radiative
cooling, AGN feedback, and mixing of high and low entropy gas due to
merging activity. Disentangling the effects of these processes
will require a concerted effort from both observations and
simulations. 

\section{Summary and perspectives}

The explosion of high quality data available from the most recent
generation of X-ray satellites has allowed unprecedented insights to
be gained into cluster physics. The parallel increase in computing
power available for large hydrodynamic simulations has facilitated
increasingly sophisticated comparisons between observations and
theory. The resulting cross-fertilisation has had profoundly
beneficial effects on our understanding of the formation and evolution
of structure in the Universe.

X-ray observations have allowed strong constraints to be put on the
form of the mass density profile in clusters, suggesting a cusped form and a
variation with mass in agreement with the predictions from numerical
simulations. This result is true over the entire measurable radial
range ($\sim 0.001$--$0.8\,\rv$) in morphologically relaxed clusters,
groups, and individual galaxies. 

The observed X-ray mass-temperature ($M$--$T$) relation has been
measured using data from \xmm and \chandra with several samples of
moderate size, yielding results which are in very good agreement both
in terms of slope and normalisation. This agreement can be attributed
both to better data quality and to improved data analysis
techniques. However, there is still an offset between the observed
\mt\ normalisation and that determined from numerical
simulations. While a part of this discrepancy may come from details of
the X-ray analysis (perhaps principally the assumption of hydrostatic
equilibrium), it is also clear that since simulations currently do not
reproduce the gross scaling properties of the cluster population
(e.g., the $L_X$--$T$ relation) the physics governing the baryonic
component are incompletely understood. 

We can get an insight into this by looking at the entropy of the
ICM. The entropy-temperature relation is shallower than expected, $S
\propto T^{0.65}$, such that poor systems have higher
entropy than rich systems relative to that expected from pure
gravitational collapse.  Entropy profiles of clusters formed purely
through gravitational processes are self-similar (once scaled
appropriately) and are characterised by a radial dependence $S \propto
R^{1.1}$ outside the central regions ($R > 0.1\,\rv$), a slope
characteristic of shock heating of the ICM. The entropy profiles of
observed clusters are self-similar outside the central regions once
scaled by the empirical scaling, and display a slope similar to that
expected from shock heating. Thus the entropy is higher throughout the
ICM in poor systems. It is not yet clear what physical mechanism is
responsible, but the most obvious candidates are preheating and AGN
activity. In the central regions ($R< 0.1\,\rv$), there is a clear
breaking of similarity. This is probably due to the combined action of
radiative cooling, AGN heating and mixing of high and low entropy gas
during mergers.

Most of the above results have been derived from observations of
morphologically relaxed systems, and are thus they are neither
representative of the cluster population as a whole, nor useful from
the point of view of establishing the intrinsic scatter about the
scaling relations. The challenge now is in the observation of larger
samples with unbiased mass and redshift sampling. These should be used
to i) further test the self-similarity of form; ii) derive the exact
slope, normalisation and intrinsic scatter of the \mt\ relation,
requiring cross-calibration and combination of different mass
estimation methods including X-ray, lensing, and dynamical; iii)
compare with state of the art simulations to probe the sources and
timescales of the non-gravitational energy input which alters the
similarity of the cluster population.

\section*{Acknowledgements}

I would like to thank Sophie Maurogordato and Laurence Tresse for
organising such an excellent meeting, and the Program and Scientific
Advisory Committees for the invitation to present this review and for
partial financial support. I warmly thank my collaborators for their
contributions to some of the papers discussed in this review. This
work received partial support from a 
Marie Curie Intra-European Fellowship (Contract No. MEIF-CT-2003-500915).

\end{document}